\documentclass[10pt,aps,prl,twocolumn,superscriptaddress]{revtex4-1}
\usepackage{graphicx}
\usepackage{amssymb}
\usepackage{amsmath}
\usepackage{appendix}
\usepackage{comment}
\newcommand{\al}[1]{\begin{aligned}#1\end{aligned}}

\newcommand{\eq}[1]{\begin{equation}#1\end{equation}}

\begin{document}

\title{Exploiting Coherence in Nonlinear Spin-Superfluid Transport}

\author{Yaroslav Tserkovnyak}
\affiliation{Department of Physics and Astronomy, University of California, Los Angeles, California 90095, USA}
\author{Mathias Kl{\"a}ui}
\affiliation{Institut f{\"u}r Physik, Johannes Gutenberg-Universit{\"a}t Mainz, 55099 Mainz, Germany}

\begin{abstract}
We show how the interference between superfluid spin currents can endow spin circuits with coherent logic functionality. While the hydrodynamic aspects of the linear-response collective spin transport obviate interference features, we focus on the nonlinear regime, where the critical supercurrent is sensitive to the phase accumulated by the condensate in a loop geometry. We propose to control this phase by electrical gating, tuning the spin-condensate coherence length. The nonlinear aspects of the spin superfluidity thus naturally lend themselves to the construction of logic gates, uniquely exploiting the coherence of collective spin currents. Vice versa, this functionality can be used to reveal the fundamental properties of spin superfluids.
\end{abstract}

\maketitle

\textit{Introduction.}|Spin currents in insulators have attracted much interest due to the possibility to transmit spin angular momentum with no associated charge flow. This may ultimately eliminate Joule heating, a prevalent dissipation mechanism in electronic and spintronic devices based on charge currents. In magnetic insulators, spin currents are carried by magnons \cite{chumakNATP15}, the quanta of the collective electron-spin excitations (spin waves) in magnetically-ordered media. As spin-1 particles, a net flow of magnons yields a pure spin current, transmitting information in the form of angular momentum. Unfolding a range of basic transport phenomena as well as considerable application potential, the investigation of generation and detection of pure magnonic spin currents in insulators has garnered significant attention. Spin currents generated by spin pumping \cite{tserkovPRL02sp}, thermal fluctuations \cite{uchidaAPL10,*bauerNATM12,*gepragsNATC16,*guoPRX16,*guoAPL16}, and electrical spin injection due to the spin Hall effect \cite{cornelissenNATP15,*goennenweinAPL15,*shanAPL17} have been studied in ferrimagnetic garnets like the insulating ferrimagnet Y$_3$Fe$_5$O$_{12}$, compensated ferrimagnets such as Gd$_3$Fe$_5$O$_{12}$, and in insulating ferrites.

Inherently, insulating magnets exhibit low damping, enabling long-distance spin propagation and thus efficient transport of spin information. The detection of magnonic spin currents is typically achieved by means of the inverse spin Hall effect \cite{sinovaRMP15} in an adjacent heavy-metal layer. At present, the magnonic currents generated by spin injection are conventionally diffusive in nature \cite{cornelissenNATP15}, exhibiting incoherent propagation and an exponential decay with increasing distance.

On the applications-related side, it was shown that magnon-based logic operations can be realized in structures employing yttrium iron garnet as a spin conduit. Incoherent magnons have been used in Ref.~\cite{ganzhornAPL16}, based on the addition of the diffusive spin signals. To fully exploit the wave nature of magnons, however, phase coherence has to be used to allow for interference effects. In particular, complex functions like majority gates, which conventionally require many semiconductor transistors, can be implemented easily using phase-coherent magnons \cite{klinglerAPL14,*fischerAPL17}. A coherent spin-wave bus thus enables the implementation of fully-functional superposition-based magnonic logic, highlighting the potential of this new information-processing approach. However, so far, the necessary coherent magnons have been generated using microwave excitations with antennas~\cite{chumakNATP15}, an approach that does not scale and is not naturally compatible with the desired integrated logic processors.

To become practical, one needs to realize dc generation of coherent magnonic spin currents, as only this allows one to fully exploit the power of coherent spin transport. To this end, we study the interference of multiple coherent collective spin currents. We find that while the hydrodynamic aspects of the spin superflow preclude interference in linear response, efficient interference effects are found in the nonlinear regime. Specifically, by exploiting a loop geometry with two coherent spin-current branches, we investigate the role of the interference in determining the critical spin-superfluid transmission. Finally, we suggest to use this result to implement logic functions, such as a functionally complete NAND gate.

\begin{figure}[!t]
\includegraphics[width=0.8\linewidth]{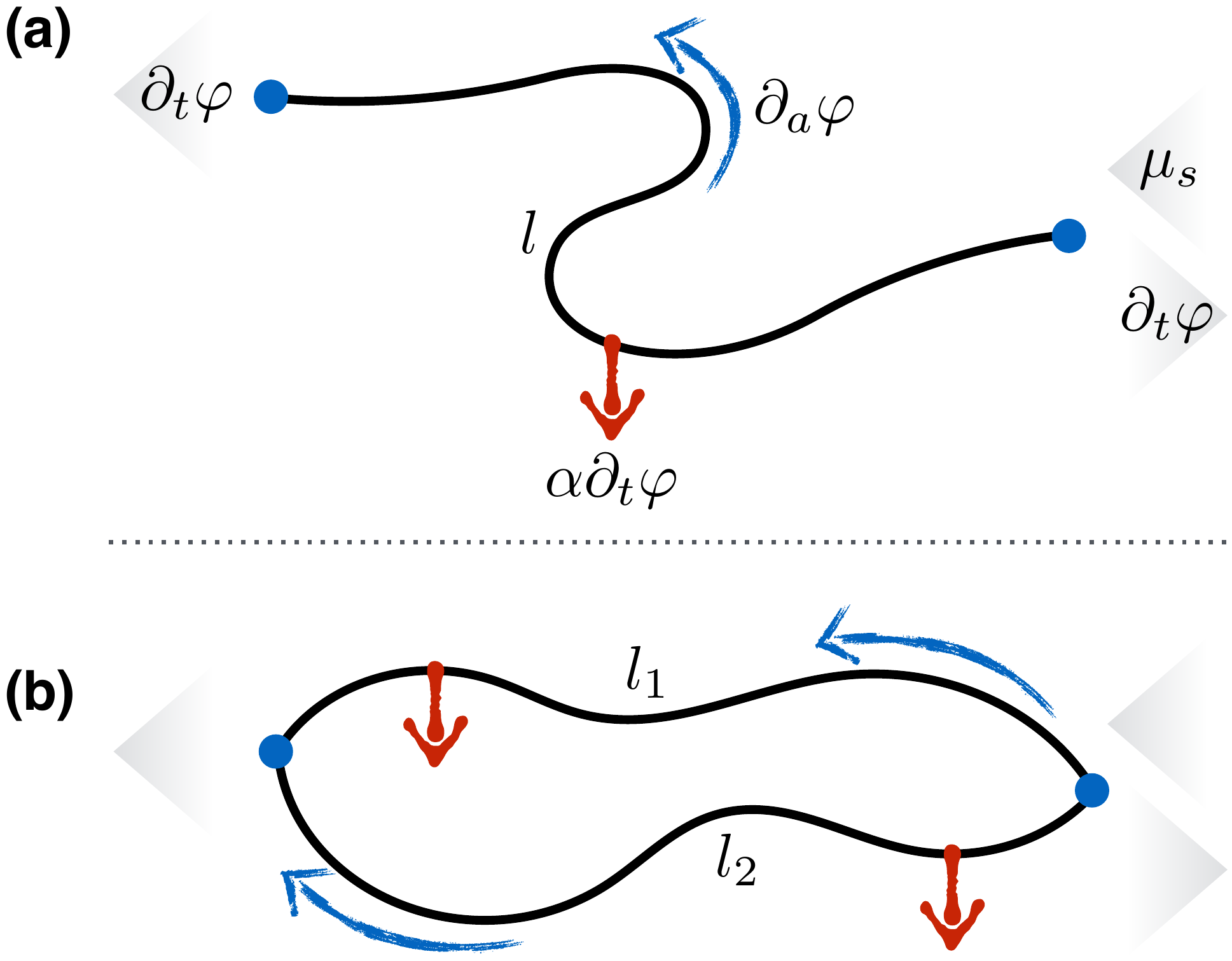}
\caption{Hydrodynamic spin transport in the linear response. (a) Single channel of length $l$ collectively transmitting spin current $\propto\partial_a\varphi$, which is injected on the right at the rate $\propto\mu_s$. Pumping $\propto\partial_t\varphi$ ejects spin currents at both ends by spin pumping, and along the length of the conduits the Gilbert damping $\alpha$ leads to an attenuation. (b) Superposition of two similar spin flows in a loop geometry composed of two branches of lengths $l_{1(2)}$. $
\varphi$ here is the azimuthal angle of easy-plane magnetic dynamics.}
\label{hydro}
\end{figure}

\textit{Spin superfluidity in linear response.}|In Refs.~\onlinecite{takeiPRL14,*takeiPRB14}, a collective spin current polarized along the $z$ axis and transmitted via the easy-$xy$-plane magnetic dynamics \cite{soninJETP78,*soninAP10} was proposed to be injected (detected) using the (inverse) spin Hall effect \cite{sinovaRMP15}. The associated spin current, $j_s=-A\partial_a\varphi$ [in the quasi-one-dimensional (1D) geometry parametrized by $a$; see Fig.~\ref{hydro}(a)], mimics closely the mass flow in a neutral superfluid \cite{halperinPR69}, while the boundary conditions $j_s=g(\mu_s-\partial_t\varphi)$, which reflect the injection and detection of spin at the channel's ends, are akin to the Andreev reflection at the normal-metal$\mid$superconductor interfaces \cite{nazarovBOOK09}. $\varphi$ here is the precession angle of the magnetic order parameter in the easy plane, $A$ is the order-parameter stiffness, $\mu_s$ is the (spin Hall-induced) spin accumulation (polarized along the $z$ axis) in the normal-metal contacts, and we are assuming a linear response (and thus only a small tilting of the order parameter out of the $xy$ plane). A crucial property in magnetic materials is the Gilbert damping, which, in this regime, sinks the angular momentum at a rate of $\alpha\partial_t\varphi$, per unit length, governed by the (dimensionless) Gilbert-damping constant $\alpha$.

In a steady state established in response to a dc bias $\mu_s$, the frequency $\partial_t\varphi\equiv\omega$ must be uniform along the full length of the channel. Balancing the spin flow at the boundaries (assuming the same spin conductances $g$) with the net Gilbert-damping loss $\alpha\omega l$, we obtain $\omega=\mu_s/(2+\alpha l/g)$. In the loop geometry of Fig.~\ref{hydro}(b), where one may anticipate interference features, the steady-state frequency is instead given by a similar expression as above, only replacing $l\to\tilde{l}=l_1+l_2$, i.e., with the total circumference of the circuit. Since the output spin current is given by $g\omega$, it does not depend on the spin texture $\partial_a\varphi$ and the associated stiffness $A$ for low excitations.

As the input bias $\mu_s$ is increased and the order-parameter winding $\partial_a\varphi$ is progressively stepped up in response, however, it will develop inhomogeneously along the loop branches. While it has no consequence for the transmitted signal in the linear response, it will have an effect on the Landau-like criterion for the superflow stability \cite{soninJETP78}. In particular, we may anticipate a larger critical current to correspond to a more uniform distribution of the flow along the two branches in the geometry of Fig.~\ref{hydro}(b). This condition, in turn, is sensitive to the interference of the two spin supercurrents, which can be controlled by the relative lengths of the two branches, in units of the respective coherence lengths. The nonlinear spin transport through the multiply-connected circuits can thus be controlled geometrically as well as by gating relevant magnetic properties along the lengths of the spin conduits. This will provide the basis for logic functionality as detailed later.

We now proceed to study collective nonlinear dynamics and spin transport in a (ferro)magnetic insulator, within the Landau-Lifshitz-Gilbert (LLG) phenomenology \cite{landauBOOKv9,*gilbertIEEEM04} for bulk dynamics and the spin Hall phenomenology \cite{tserkovPRB14} for the spin injection and detection at the boundaries. After briefly summarizing the pertinent equations and optimizing the notation, we will study the stability of spin superflow in the geometries of Fig.~\ref{hydro}, with a focus on the loop geometry that will yield interference and thus lay the foundations for the desired logic functionality.

\textit{LLG theory of the nonlinear spin transport.}|The (nonlinear) LLG dynamics in the (insulating) bulk,
\eq{
s(1+\alpha\mathbf{n}\times)\dot{\mathbf{n}}=\delta_\mathbf{n}F\times\mathbf{n}+\boldsymbol{\tau}\,,
\label{LLG}}
is constructed in terms of the free-energy functional
\eq{
F[\mathbf{n}]=\int d^3r\left[A(\partial_i\mathbf{n})^2+Kn_z^2\right]/2\,.
\label{F}}
$\boldsymbol{\tau}$ here stands for any applied spin torques, $s$ is the equilibrium spin density, and $K>0$ is the superfluidity-stabilizing \cite{soninJETP78} easy-plane anisotropy. The order parameter undergoes directional dynamics constrained by $|\mathbf{n}|\equiv1$. We can rewrite Eq.~\eqref{LLG} as a hydrodynamic continuity equation:
\eq{
s(1+\alpha\mathbf{n}\times)\dot{\mathbf{n}}=-\partial_i\mathbf{j}_i+Kn_z\mathbf{z}\times\mathbf{n}+\boldsymbol{\tau}\,,
}
where $\mathbf{j}_i\equiv-A\mathbf{n}\times\partial_i\mathbf{n}$ is recognized to be the spin flow in the $i$th direction.

For the boundary conditions, attaching a heavy metal with the interface area $S$ and normal $\mathbf{k}$ results in the spin-injection current density (i.e., torque per unit area)
\eq{
\mathbf{j}_s=\mathbf{j}^{\rm (SH)}_s-\mathbf{j}^{\rm (pump)}_s=\vartheta\mathbf{n}\times(\mathbf{k}\times\mathbf{j})\times\mathbf{n}-g\mathbf{n}\times\dot{\mathbf{n}}\to\frac{\delta\boldsymbol{\tau}}{\delta S}\,,
}
where $\mathbf{j}$ is the electrical current density applied to the metal. $\vartheta\equiv(\hbar/2e)\tan\theta_{\rm SH}$, in terms of the effective spin Hall angle $\theta_{\rm SH}$, and $g\equiv(\hbar/4\pi)g^{\uparrow\downarrow}$, in terms of the effective spin-mixing conductance (per unit area) $g^{\uparrow\downarrow}$, both including the interplays of the spin Hall and spin-pumping injection, reflection, and backflow of electron spins in the metal. We are keeping here only the leading-order in spin-orbit interaction effects \cite{tserkovPRB14}. We will henceforth set $\mathbf{k}\to\mathbf{x}$ and $\mathbf{j}\to j\mathbf{y}$, so that $\mathbf{k}\times\mathbf{j}\to j\mathbf{z}$. The same metal can be used for detecting magnetic dynamics, according to the Onsager-reciprocal spin-motive force \cite{tserkovPRB14}:
\eq{
\boldsymbol{\epsilon}=\vartheta(\mathbf{n}\times\dot{\mathbf{n}})\times\mathbf{k}=\vartheta\mathbf{j}^{\rm (pump)}_s\times\mathbf{k}/g\,,
\label{smf}}
which, in a closed circuit, would induce a current density $\mathbf{j}=\sigma\boldsymbol{\epsilon}/d$, where $\sigma$ is the metal film's conductivity and $d$ its thickness.

Let us parametrize $\mathbf{n}(\theta,\varphi)$ by the polar angle $\theta$ and the azimuthal angle $\varphi$. Let $(\mathbf{n},\boldsymbol{\theta},\boldsymbol{\varphi})$ be the local (right-handed) coordinate system, such that
\eq{
\partial_i\mathbf{n}=\boldsymbol{\theta}\partial_i\theta+\boldsymbol{\varphi}\partial_i\varphi\sin\theta\,.
}
It then follows that
\eq{\al{
\partial_i(\mathbf{n}\times\partial_i\mathbf{n})=&-\boldsymbol{\theta}\frac{\partial_i(\partial_i\varphi\sin^2\theta)}{\sin\theta}\\
&+\boldsymbol{\varphi}\left[\partial_i^2\theta-\frac{1}{2}(\partial_i\varphi)^2\sin2\theta\right]\,.
}}
Projecting the LLG equation \eqref{LLG} in the bulk on $\boldsymbol{\theta}$ and $\boldsymbol{\varphi}$, we respectively get
\eq{
s(\dot{\theta}-\alpha\dot{\varphi}\sin\theta)=-A\frac{\partial_i(\partial_i\varphi\sin^2\theta)}{\sin\theta}
\label{bulk1}}
and
\eq{
s(\dot{\varphi}\sin\theta+\alpha\dot{\theta})=A\left[\partial^2_i\theta-\frac{(\partial_i\varphi)^2}{2}\sin2\theta\right]+\frac{K}{2}\sin2\theta\,.
\label{bulk2}}
Switching to the natural units for the problem, we measure $\partial_t$ in units of $K/s$ and $\partial_i$ in units of $\sqrt{K/A}$ (the magnetic speed of sound then becomes $c=\sqrt{KA}/s\to1$). The bulk equations of motion then become dimensionless as $s$, $A$, and $K$ drop out.

\textit{Critical superflow in a single conduit.}|In a 1D superfluid channel of length $l$, whose position is parametrized by $a$, the bulk equations \eqref{bulk1} and \eqref{bulk2} reduce to
\eq{\al{
\dot{\theta}-\alpha\dot{\varphi}\sin\theta&=-\frac{\partial_a(\partial_a\varphi\sin^2\theta)}{\sin\theta}\,,\\
\dot{\varphi}\sin\theta+\alpha\dot{\theta}&=\left[\partial^2_a\theta+\frac{1-(\partial_a\varphi)^2}{2}\sin2\theta\right]\,.
\label{bulk}}}
Placing the normal metals at the two ends ($a=0$ and $l$), the boundary conditions projected onto $\boldsymbol{\theta}$ result in
\eq{
a=0,l:~~~(\mp\partial_a\varphi+\mathfrak{g}\dot{\varphi}-\mathfrak{j})\sin\theta=0\,,
\label{bc1}}
and for $\boldsymbol{\varphi}$:
\eq{
a=0,l:~~~\partial_a\theta\mp\mathfrak{g}\dot{\theta}=0\,.
\label{bc2}}
Here, the dimensionless constants $\mathfrak{g}\equiv(g/s)\sqrt{K/A}$ and $\mathfrak{j}\equiv\vartheta j/\sqrt{AK}$ (which may be different at the two ends) parametrize the strengths of the spin pumping and the spin Hall torques at the interfaces. They both may include the geometric enhancement factor $S/S_m$ (where $S_m$ is the magnetic wire cross section), which we are omitting for simplicity. However we note that analogous to a hydrodynamic description using a tapered geometry, potentially one can enhance the spin current density by this geometrical factor. We are supposing that the metal contacts are on top of the magnet with the same normal $\mathbf{k}$ (on the bottom, the relative sign in front of $j$ would flip, as in our original Ref.~\cite{takeiPRL14}). Let us note that $\theta\equiv0$ is a good solution (albeit possibly unstable) of Eqs.~\eqref{bulk}-\eqref{bc2}, as all the spin torques and currents vanish in this trivial case.

In a stable dynamic steady state, we can set $\dot{\theta}(a,t)\equiv0$ and $\dot{\varphi}(a,t)\equiv\omega$ (constant). Defining $v\equiv-\partial_a\varphi$ (corresponding to the velocity of the superfluid condensate), we rewrite the above equations as
\eq{\al{
-\alpha\omega\sin^2\theta&=\partial_a(v\sin^2\theta)\,,\\
\omega\sin\theta&=\left[\partial^2_a\theta+(1-v^2)\sin\theta\cos\theta\right]\,,
\label{bulks}}}
with the boundary conditions (supposing $\theta\neq0$)
\eq{
a=0,l:~~~\pm v+\mathfrak{g}\omega-\mathfrak{j}=0\,,\,\,\,\partial_a\theta=0\,.
\label{bc}}
Note that $v\to1$ corresponds to the Landau criterion, according to which a static  spiral becomes energetically unstable \cite{soninJETP78}. We can see this from the energy density in Eq.~\eqref{F}, which is $\propto(1-v^2)n_z^2$, in our units: At $v>1$, the uniform out-of-plane state $n_z\equiv1$ has the lowest energy.

Let us start by looking for solutions with a constant $\theta\neq0$. From Eqs.~\eqref{bulks}, we then get:
\eq{
-\alpha\omega=\partial_av~~~{\rm and}~~~\omega=(1-v^2)\cos\theta\,.
}
It is clear that a constant-$\theta$ solution implies also a constant $v$, which requires that either $\alpha$ or $\omega$ vanish. $\omega\to0$, furthermore, necessitates $\mathfrak{j}\equiv\mathfrak{j}(0)=-\mathfrak{j}(l)$. In this case, $v=\mathfrak{j}$ carries the spin Hall-injected spin current between the contacts without any dissipation. $\theta=\pi/2$ up to $\mathfrak{j}\to1$, at which point there is a first-order phase transition to $\theta=0$, for $\mathfrak{j}>1$. Setting $\alpha=0$ would generally result in constant-$\theta$ solutions. Supposing $\mathfrak{g}$ entering Eqs.~\eqref{bc} is the same at both ends,
\eq{
v=\frac{1-p}{2}\mathfrak{j}\,,\,\,\,\omega=\frac{1+p}{2\mathfrak{g}}\mathfrak{j}\,,\,\,\,{\rm and}\,\,\,\cos\theta=\frac{\omega}{1-v^2}\,,
\label{vw}}
where $\mathfrak{j}(0)=\mathfrak{j}$ and $\mathfrak{j}(l)=p\mathfrak{j}$, with $p$ parametrizing the injection polarization. In the antisymmetric case, $p=-1$, we reproduce the above finite-$v$, zero-$\omega$ solution (since in the absence of dynamics, the Gilbert damping is inconsequential). In the symmetric case, $p=1$, a finite-$\theta$ solution (with finite $\omega$ and zero $v$) exists up to the critical bias $\mathfrak{j}_c=\mathfrak{g}$. For an arbitrary $p$, the critical bias $\mathfrak{j}_c$ is reached when $\omega=1-v^2$. We can see that $\mathfrak{j}_c\leq2/(1-p)$ and $2\mathfrak{g}/(1+p)$, corresponding respectively to $v,\omega\leq1$.

\begin{figure}[!t]
\includegraphics[width=0.9\linewidth]{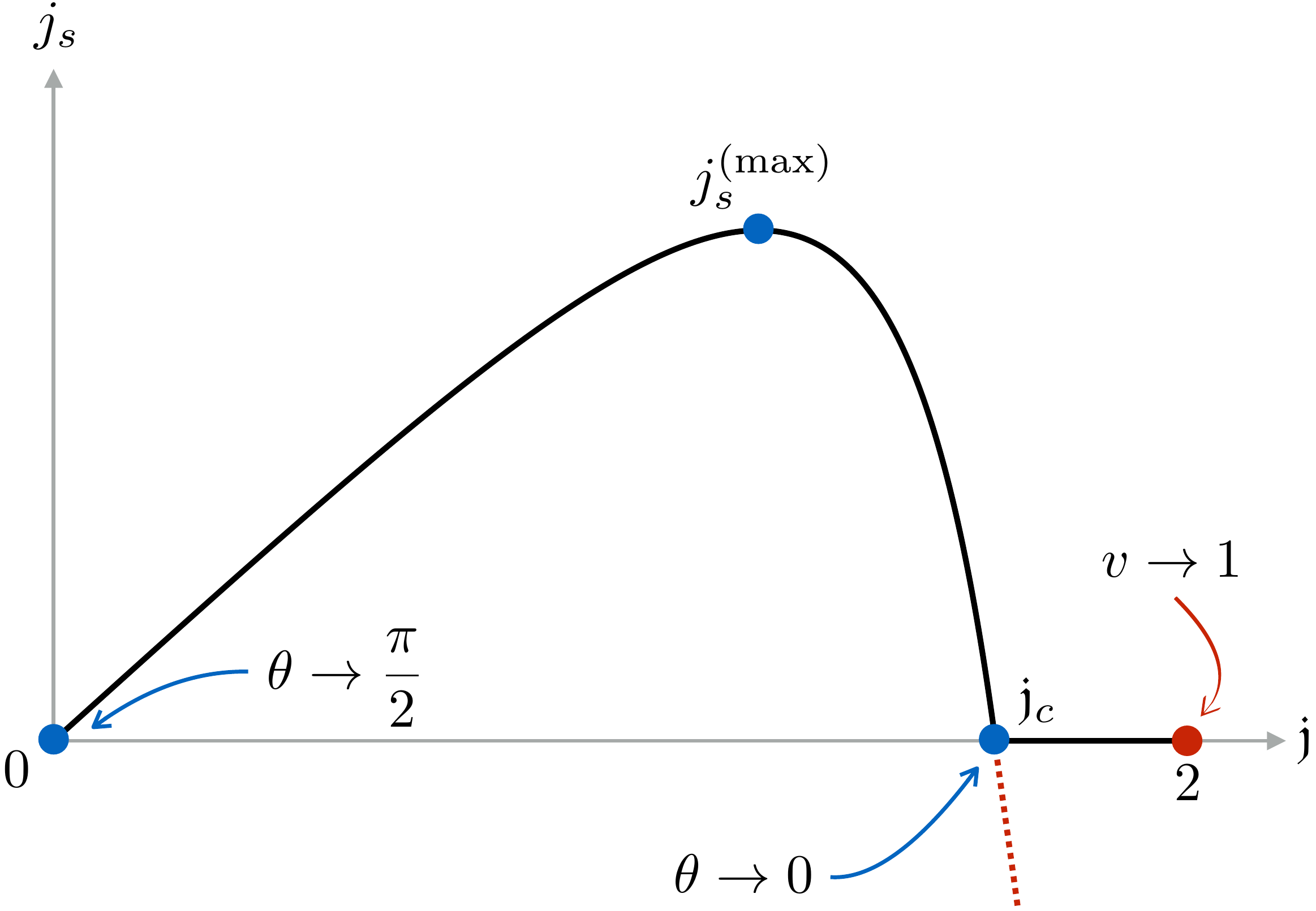}
\caption{Spin current \eqref{tjs}, which governs the detected motive force $\epsilon=\vartheta j_s/g$, in the case of $p=0$ and choosing $\mathfrak{g}=2$. Note that the picture would simply flip for the opposite bias, $\mathfrak{j}<0$.}
\label{js}
\end{figure}

When $p\neq-1$, the steady-state solutions are dynamic and the critical angle $\theta\to0$ is reached in a second-order fashion (cf. Fig.~\ref{js}). The transmitted ($\mathbf{z}$-polarized) spin-current density, in this case,
\eq{
j_s=sv(1-\cos^2\theta)\,,
\label{tjs}}
is maximized at some intermediate bias, between $0$ and $\mathfrak{j}_c$ [i.e., the critical point where $\theta$ vanishes; note that both $v$ and $\theta$ here depend on $j$ according to Eqs.~\eqref{vw}]. This means that one can maximize the injected spin current density by choosing the appropriate injector current. In the special case when $p=0$ (corresponding to the injection at $a=0$), the transmitted spin current,
\eq{
j_s=s\mathfrak{j}(1-\cos^2\theta)/2\,,
}
will result in the ($\mathbf{y}$-oriented) motive force \eqref{smf} $\epsilon=\vartheta j_s/g$.

\textit{Nonlinear superflow interference.}|Having established how to maximize the single spin currents, the next step is to study the interplay of multiple spin currents as a prerequisite for using them for logic. We now study in particular the case of two (interfering) superfluid channels connected at the ends. Representing them as a circle, we start with the simplest case of two metal contacts: injector at $a=0$ and detector at $a=l'$, with the full loop length given by $l$. Barring superfluid phase slips \cite{kimPRB16,*kimPRL16}, we restrict $\theta$ to the interval of $(0,\pi)$. (In other words, the spin texture is not allowed to sweep over the south or north poles.) This allows us to define the topological invariant
\eq{
2\pi n=\int_0^l da\,v\,,
\label{tc}}
corresponding to the net (azimuthal-angle) winding number $n\in\mathbb{Z}$ of the order-parameter texture placed on the circle.

\begin{figure}[!t]
\includegraphics[width=0.8\linewidth]{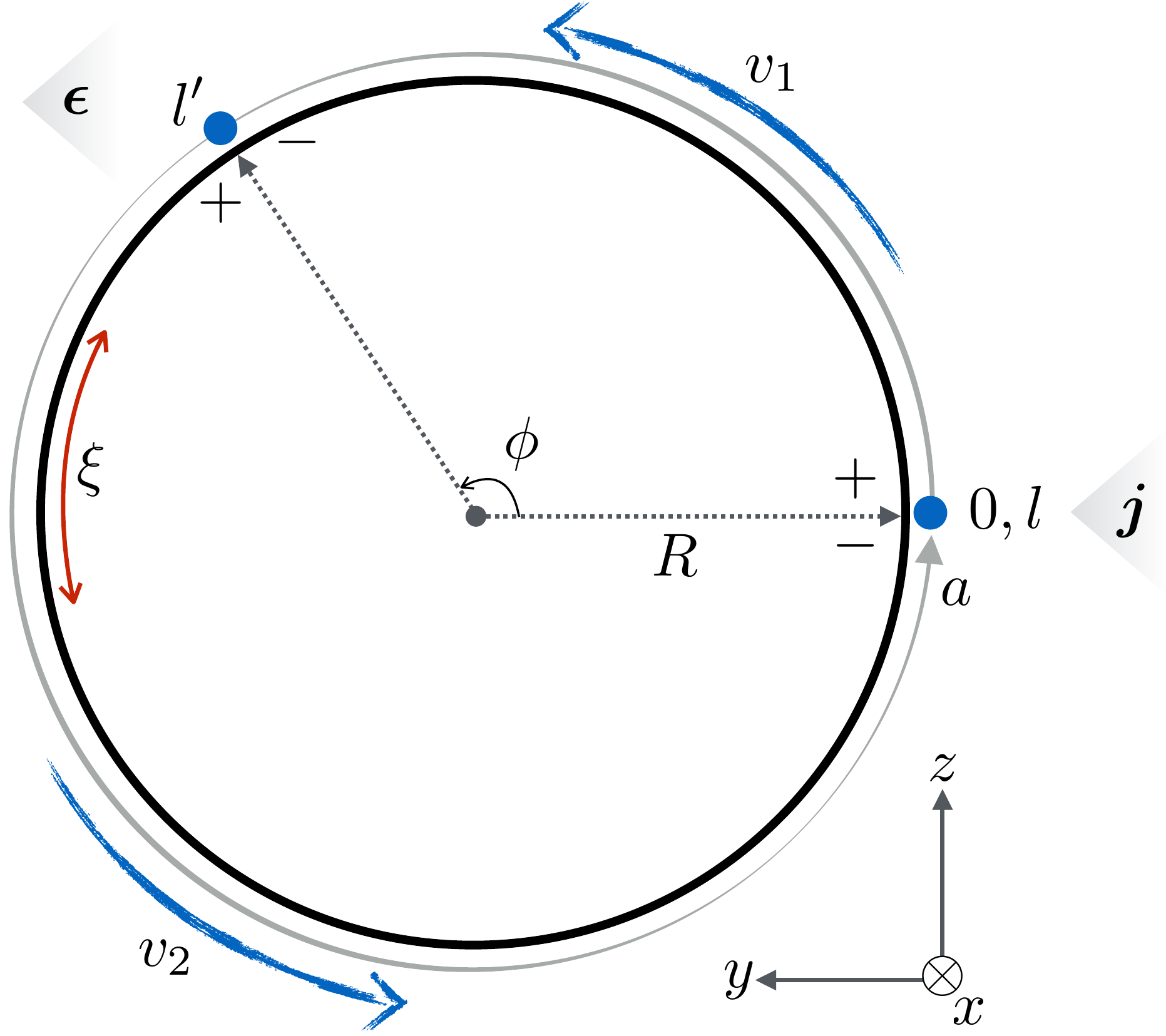}
\caption{Schematic of the circular configuration to exhibit interference of two (nonlinear) spin superfluids. The critical current is maximized at special relative angles $\phi$ between the injector and detector leads, which are determined by $\xi/R$ according to Eq.~\eqref{dl}.}
\label{sch}
\end{figure}

We are looking for steady-state solutions of the same bulk Eq.~\eqref{bulks}, adjusting the boundary conditions as
\eq{
a=0,l':~~~\left.v\right|^+_-+\mathfrak{g}\omega-\mathfrak{j}=0\,,\,\,\,\left.\partial_a\theta\right|^+_-=0\,.
\label{loop}}
See Fig.~\ref{sch} for a schematic explaining the geometry and notation. In the absence of damping, $\alpha\to0$, and for subcritical driving, let us try to look for the superfluid velocities that are uniform in the two sections, given by $v_1$ and $v_2$, while the polar angle $\theta$ is the same throughout. Setting $\mathfrak{j}(0)=\mathfrak{j}$ and $\mathfrak{j}(l')=0$,
\eq{
v_2-v_1=\mathfrak{g}\omega-\mathfrak{j}~~~{\rm and}~~~v_1-v_2=\mathfrak{g}\omega\,,
}
subject to the topological constraint \eqref{tc}: $v_1 l'+v_2(l-l')=2\pi n$. We thus find:
\eq{
v_1-v_2=\frac{\mathfrak{j}}{2}~~~{\rm and}~~~\frac{v_1+v_2}{2}=\frac{\mathfrak{j}}{2}\left(\frac{1}{2}-\frac{l'}{l}\right)+\frac{2\pi n}{l}\,.
\label{vv}}
The frequency $\omega=\mathfrak{j}/2\mathfrak{g}$ (which governs the detected motive force) is $l'$ independent. Note that the frequency $\omega=(1-v^2)\cos\theta$ can generally not be the same for a common angle $\theta$ in the two sections. This means that the above steady-state solution would be valid only in the linear-response regime. In the general nonlinear case, $\theta(a)$ must necessarily develop inhomogeneities, with the exception of the special scenarios that yield $|v_1|=|v_2|$ according to Eq.~\eqref{vv}.

We could initialize a uniform state with $n=0$, in the absence of a bias, followed by ramping up the current $\mathfrak{j}$. If $l'\neq l/2$, the two branches will transmit the input current asymmetrically, so that a critical current would be reached in one of them before the other. The texture can then undergo a phase slip to a different winding number $n$, depending on the ratio $l'/l$, with a possibility to reach a steady state with a higher critical current. The symmetrical (i.e., nonfrustrated) case, $l'=l/2$, corresponds to the highest critical current $\mathfrak{j_c}$, when $n=0$, so that $v_1=-v_2=\mathfrak{j}/4\equiv v$. As before, $\mathfrak{j_c}$ is found from $\omega=\mathfrak{j}/2\mathfrak{g}=1-v^2$. If $\mathfrak{g}\gg1$, in particular, the critical current is obtained from $v\to1$ and is thus twice the result for a single 1D channel (with $p$ set to $0$). In order to maintain the symmetrical superfluid flow, $v_1=-v_2$, allowing us to reach the highest critical current, we obtain the following condition from Eqs.~\eqref{vv}:
\eq{
\frac{\mathfrak{j_c}}{2}\left(\frac{l'}{l}-\frac{1}{2}\right)=\frac{2\pi n}{l}\,.
}
We thus obtain the maximal superflow at $l'=l/2$ (and $n=0$) as well as at the positional increments of
\eq{
\Delta l'=4\pi n/\mathfrak{j}_c\,.
\label{dl}}
Note that, restoring the physical units , $\mathfrak{j}_c\sim\xi^{-1}$, where $\xi\equiv\sqrt{A/K}$ is the magnetic healing length. The size of the ring thus has to be larger than but comparable to this length scale, for the optimal geometric characteristics and sensitivity.

If the current is injected symmetrically at both contacts, $\mathfrak{j}(0)=\mathfrak{j}(l')=\mathfrak{j}$, we find, according to Eqs.~\eqref{loop}: $v_1=v_2=2\pi n/l\equiv v$ and $\omega=\mathfrak{j}/\mathfrak{g}$. The common polar angle and stability considerations are then derived from $\omega=(1-v^2)\cos\theta$, as in the single-conduit case. This results in an $n$-dependent critical current. In particular, since $n=0$ corresponds to the highest current, the $n=0$ configuration can be initialized by driving a symmetric bias that is subcritical to this state only.

\textit{Discussion.}|The nonlinear interference physics affecting critical spin current in the multiply-connected geometries (e.g., a two-branch loop shown in Fig.~\ref{sch}) is a key result of our study. The natural unit of length that controls this interference is set by the coherence length $\xi$ [cf. Eq.~\eqref{dl}], so that we may expect the strongest interference effects on the critical current to be for $l$ greater than $\xi$. While in our discussion pertaining to Fig.~\ref{sch}, we control the relative phase between the two superfluid branches by sliding the position $l'$ along the loop, this is not practical in a useful device. A more apt approach is to locally vary $A$ and/or $K$  (and thus $\xi$) in one of the branches. This can be achieved, for instance, by electrostatic gating \cite{weisheitSCI07,*maruyamaNATN09} or by locally applying strain \cite{leiNATC13,*finizioPRAP14}, which can manipulate anisotropies and other magnetic properties. We can thus enable or disable an effective transmission of a large input signal, by, respectively, lowering or raising the value of the critical current, depending on the exact structure. An additional potential functionality may also be achieved by an appropriate topological initialization to some desired winding $n$. In terms of the logic gates, this can be used to accomplish the AND and NOT gates, which together provide a functionally complete set that allows for the implementation of any logic function.

Note that numerically exploring the nonlinear spin dynamics in the supercritical regime, particularly with an eye on tunable steady-state self-oscillations, is a potentially interesting avenue of research. On another front, the (heretofore disregarded) thermally-induced phase slips \cite{kimPRB16} may offer an alternative mechanism for how the interplay of nonlinearities and interference along multiple spin-superfluid branches can be exploited.

\begin{acknowledgments}
We are grateful to Benedetta Flebus for helpful discussions. Work by YT was supported by the U.S. Department of Energy, Office of Basic Energy Sciences under Award No.~DE-SC0012190. Work by MK was supported by the German Research Foundation (SFB TRR173 Spin+X, SPP 1538 Spin Caloric Transport) as well as the EU (INSPIN FP7-ICT-2013-X 612759) and the State Centre for Interdisciplinary and Emerging Materials (CINEMA). The collaboration, furthermore, benefited from the conference on ``Spin Coherence, Condensation, and Superfluidity" supported by the Army Research Office under Contract No. W911NF-17-1-0106.
\end{acknowledgments}

\end{document}